\documentclass[aps,pre,reprint,groupedaddress,superscriptaddress,amsmath,amssymb,showpacs]{revtex4-2}
\usepackage{graphicx}
\usepackage{bm}
\usepackage{physics} 

\begin{document}

\title{Electric field in spatially inhomogeneous non-neutral plasma }

\author{S. Ya. Bronin}
\affiliation{Joint Institute for High Temperatures of the Russian Academy of Sciences, Izhorskaya St. 13, Bldg. 2, Moscow 125412, Russia}
\author{E. V. Vikhrov}
\affiliation{Joint Institute for High Temperatures of the Russian Academy of Sciences, Izhorskaya St. 13, Bldg. 2, Moscow 125412, Russia}
\affiliation{Keldysh Institute of Applied Mathematics of the Russian Academy of Sciences, Miusskaya sq., 4, Moscow, 125047, Russia}
\author{B. B. Zelener}
\affiliation{Joint Institute for High Temperatures of the Russian Academy of Sciences, Izhorskaya St. 13, Bldg. 2, Moscow 125412, Russia}
\author{B. V. Zelener}
\email{bzelener@mail.ru}
\affiliation{Joint Institute for High Temperatures of the Russian Academy of Sciences, Izhorskaya St. 13, Bldg. 2, Moscow 125412, Russia}
\date{\today}

\begin{abstract}
    We present a general expression for the probability distribution function of electric field in a plasma cloud formed by the impact of a laser pulse on a gas or a solid body. We also present the results of
    numerical calculation of this function for the case of non-interacting particles depending on the plasma cloud size. It takes into account the ionic microfield and the macrofield arising from the charge
    imbalance. As the charge imbalance increases, the effect of a sharp increase in the distribution function for large field values is observed. Good agreement between the calculation of the shift of the spectral
    line and the experiment is obtained. The results obtained are of crucial importance for diagnosing plasma in various applications.
\end{abstract}



\maketitle

    One of the methods of plasma diagnostics is the analysis of the influence of its electric field on the shape of the spectral lines of atoms and ions (the Stark effect).
In this regard, the main parameters that determine the said effect are the concentration of charges, their temperature, and the charge number~\cite{B_Griem}.

    The distribution function of a low-frequency ionic microfield in an ideal homogeneous plasma was obtained by Holtsmark~\cite{B_Griem}.
Attempts to take into account correlation effects between particles were made in~\cite{B_Baranger, B_Mozer, B_Hooper_1, B_Hooper_2, B_O_Brien, B_Kurilenkov, B_Iglesias, B_Potekhin, B_Nersisyan, B_Sadykova} (see also 
reviews ~\cite{B_Demura, B_Lisitsa}). In our papers \cite{B_Bobrov, B_Bronin}, we used the molecular dynamics method in order to calculate the distribution functions of the ionic microfield for a two-component 
singly and multiply ionized unbounded homogeneous plasma at neutral and positively charged points depending on the strong coupling parameter. In this regard, the model of ultracold plasma (UCP) was used for 
the calculation, in which the interaction between particles was described by the Coulomb law without any restrictions.

    At the same time, it is of great interest to study the distribution of electric field for systems having a finite number of particles in a limited volume, such as clusters of charged particles.
Such clusters are formed, for instance, in experiments involving the impact of femtosecond- to nanosecond laser pulses on a gas jet expanding into a vacuum or on solid targets (see, for instance~\cite{B_Sack, B_Mora_1} 
and references in them). In both cases, this leads to the formation of a plasma cloud. The plasma formed then expands into the environment. At a time greatly exceeding the duration of the laser pulse, a rather complex 
object arises, consisting of a mixture of neutral particles and expanding ions and electrons. For such clusters, the distribution of the ionic microfield was considered in~\cite{B_Romanovsky, B_Ebeling_1, B_Ebeling_2}.
The number of ions in these clusters is $<10000$. In the aforementioned papers were made estimates for the value of the microfield in the Holtsmark approximation at the center of the cluster, in which the ion density is 
distributed according to the Gauss or Levy law.

    A considerably simpler object arises when ultra-cold plasma (UCP) is formed in magneto-optical traps.

    The initial density of charged particles in the cloud is distributed according to the Gauss law, and their number can reach $10^8$.
The influence of neutral atoms can be neglected. In our papers~\cite{B_Vikhrov_1, B_Vikhrov_2, B_Vikhrov_3}  it was shown that the UCP formed in a MOT by means of a pulsed laser scatters due to the appearance of a 
supersonic ion wave. In this regard, the ion density distribution during expansion differs from the Gaussian distribution. The appearance of the wave is due to the charge imbalance resulting from the evaporation of a 
part of the electrons from the plasma cloud. All these processes take place in the clusters considered in~\cite{B_Romanovsky, B_Ebeling_1, B_Ebeling_2} as well. In a non-neutral plasma, in contrast to a neutral plasma, 
an electric macrofield arises, which can, along with the microfield, affect the spectral lines of atoms and ions.

    In experiments involving the UCP~\cite{B_Dutta, B_Feldbaum, B_Park}, macro- and microfields were studied on the basis of the influence of the Stark effect on Rb atoms excited to the Rydberg state.
In~\cite{B_Dutta, B_Feldbaum}, for instance, the ionic microfield was measured, which was formed by an ionic cloud and which affected the spectral lines of the Rb atomin $ns$ and $nd$ states at $n=20-50$.
In~\cite{B_Park}, an attempt was made to estimate the effect of the ionic microfield and macrofield on the measured spectral lines of the $42s_{1/2}-42p_{1/2}$ transition of an excited rubidium atom as a function of the 
expansion time of the rubidium UCP cloud.

In this paper, we propose a general approach for taking into account the influence of an ionic microfield and macrofield in a spatially inhomogeneous non-neutral plasma on the spectral lines of Ry-atoms.
A general expression is given for the probability distribution function of the electric field, with account taken of the micro- and macrofield. An analytical expression is obtained for the probability function of the ionic
microfield for an ideal plasma in the case of the distribution of the ion density according to the Gaussian law, depending on the radius of the plasma cloud. Analytical expressions are obtained for small radii of the plasma
cloud. Numerical calculation of the distribution of the ionic microfield as a function of the radius for the case of non-interacting particles are carried out. The possibility of taking into account the influence of the 
interaction between charges on the distribution of the ionic microfield in a bounded plasma is discussed. For the UCP of rubidium under the experimental conditions of~\cite{B_Park}, the absorption coefficient is 
numerically calculated with account taken of the micro- and macrofields.
\begin{figure}[ht!]
    \includegraphics[width=1\columnwidth]{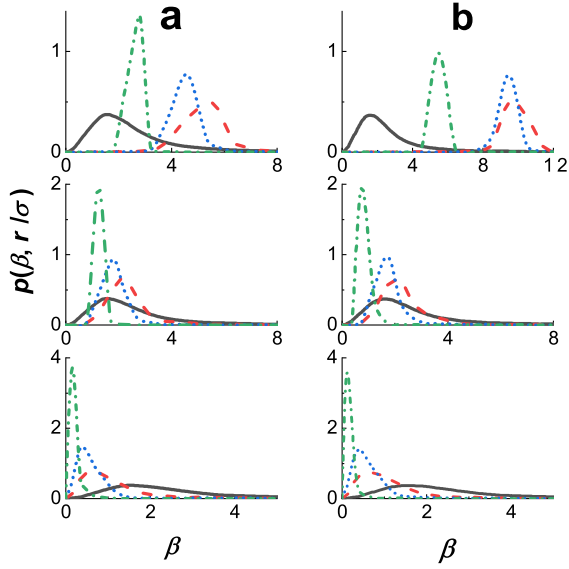}
    \caption{
                \label{Fig_1}Distribution functions $p\left( \beta, r/\sigma \right)$ calculated in accordance with~(\ref{Eq_5}),~(\ref{Eq_6}) and~(\ref{Eq_7}) for two $N_i$ values: $1000$ (a) and $10000$ (b) 
                and for four values of $r/\sigma$: $0$ corresponds to black solid line, $1$ corresponds to red dotted line, $2$ corresponds to blue dotted line, and $3$ corresponds to green dash-dotted line. 
                From top to bottom: a): $N_e/N_i = 0$, $N_e/N_i = 0.6$, $N_e/N_i = 1$; b): $N_e/N_i = 0$, $N_e/N_i = 0.2$, $N_e/N_i = 1$.
            }
\end{figure}

    Evaporation of a part of the electrons during $\sim 10$~ns results in a charge imbalance $N_e < N_i$ occurring in the UCP cloud. The resulting electric field is the sum of the fields created by 
the electrons having the spatial distribution function $n_e \left( \bm r \right) = N_e f_e \left( \bm r \right)$ and by the ions having the distribution 
$n_i \left( \bm r \right) = N_i f_i \left( \bm r \right)$ ($N_e ,  N_i \gg 1$). In the case where the lifetime of the excited state exceeds the time during which 
an electron moves beyond several average inter-electron distances, the level shift is affected by the time-averaged electron field ${\bm E}_e$:
\begin{equation}
    \label{Eq_1}
    \boldsymbol{\nabla} \cdot {\bm E}_e = -4 \pi e n_e \left( \bm r \right),
\end{equation}
where $e$ is the electron charge, $N_e$, $n_e$, $N_i$, and $n_i$ are the numbers of particles and the concentration of electrons and ions, respectively.

    The field of ions, whose displacement during the lifetime of the upper level can be neglected, is equal to:
\begin{equation}
    \label{Eq_2}
    {\bm E}_i \left( \bm r ,  \bm r_k \right) = e \sum_{k = 1}^{N_i} \frac{\bm r - {\bm r}_{k}}{|\bm r - {\bm r}_{k}|^3}.
\end{equation}
Probability distribution of different electric field values is:
\begin{equation}
    \label{Eq_3}
    P \left( \bm r ,  \bm E \right) = \Big{\langle} \delta \Big{(} \bm E - {\bm E}_e \left( \bm r \right) - {\bm E}_i \left( \bm r ,  \bm r_k \right) \Big{)} \Big{\rangle}.
\end{equation}
The ion contribution includes the term ${\bm E}_i$, which is similar to ${\bm E}_e$ and which satisfies the following equation:
\begin{equation}
    \label{Eq_4}
    \boldsymbol{\nabla} \cdot {\bm E}_i = 4 \pi e n_i \left( \bm r \right),
\end{equation}
and the term $\Delta \bm E$ (microfield), which is due to the contribution of the nearest neighbors and which forms the kernel of the distribution function of the Holtsmark type: 
${\bm E}_i \left( \bm r ,  \bm r_k \right) = {\bm E}_i + \Delta \bm E$. In the case of a plasma localized in a volume $V$, the term $\Delta \bm E$ determines the shape of the distribution 
function at the internal points of the volume, with a characteristic scale of the field magnitude $\sim e \left( N_i / V \right)^{2/3}$. At the plasma periphery, the determining field is the macrofield 
${\bm E}_m = {\bm E}_e \left( \bm r \right) + {\bm E}_i \left( \bm r \right)$ having a characteristic scale $\sim e {\left( N_i - N_e \right) / V}^{2/3}$. As the distance 
increases, the distribution function takes a delta-like form having its width proportional to the solid angle $\sim \sigma^2 / r^2$ under which the plasma volume is visible from the pointrand having its center at the 
point $E = |{\bm E}_i \left( \bm r \right)|$. This follows from the estimate: 
${\bm E}_i \left( \bm r ,  \bm r_k \right) = e \bm r \sum {cos} \vartheta_i / r^3 = e N_i \bm r \left( 1 + O \left( \sigma^2 / r^2 \right) \right) / r^3$.

    The distribution function of the electric field is given by the following expression:
\begin{widetext}
    \begin{multline}
        \label{Eq_5}
        P \left( \bm r,\bm E \right) = \int \prod_{k=1}^{N_i} d\bm r_k \frac{n_i \left( \bm r_k \right)}{N_i} \delta \Big{(} \bm E - {\bm E}_e \left( \bm r \right) - {\bm E}_i \left( \bm r,\bm r_k \right) \Big{)} = \\
        \frac{1}{\left( 2 \pi \right)^3} \int d\boldsymbol \mu \cdot exp \left[ i \boldsymbol \mu \Big( \bm E - {\bm E}_e \left( \bm r \right) \Big) \right] \cdot 
        \left( 1 -  \frac{1}{N_i} \int \left( 1 - exp \left[ -i \boldsymbol \mu e \frac{\bm r - \bm r'}{|\bm r - \bm r'|^3} \right] n_i \left( \bm r' \right) d\bm r' \right) \right)^{N_i} \approx \\
        \frac{1}{\left( 2 \pi \right)^3} \int d\boldsymbol \mu \cdot exp \left[ i \boldsymbol \mu \Big( \bm E - {\bm E}_e \left( \bm r \right) \Big) \right] \cdot exp \left[ -C \left( \bm \mu,\bm r \right) \right],
    \end{multline}
\end{widetext}
where
    \begin{equation}
        \label{Eq_6}
        C \left( \bm \mu ,  \bm r \right) = \int 1 - exp \left[ -i \boldsymbol \mu e \frac{\bm r - \bm r'}{|\bm r - \bm r'|^3} \right] n_i \left( \bm r' \right) d \bm r'.
    \end{equation}
    The required distribution function of the electric field $E = |{\bm E}|$ is given by the following integral:
\begin{equation}
    \label{Eq_7}
    \widetilde{p} \left( \bm r, E \right) = E^2 \int P \left( \bm r ,  E \boldsymbol \Omega \right) d \boldsymbol \Omega.
\end{equation}

    It is important to note that in~(\ref{Eq_5}) and~(\ref{Eq_7}), the probability distribution function of the total electric field is considered, rather than the sum of the distributions of the micro- and macrofields.
If the plasma formed has its center of symmetry at the point $\bm r= \boldsymbol 0$ so that ${\bm E}_m \left( \boldsymbol 0 \right)= 0$, then the distribution $\tilde{p} \left( \boldsymbol 0 ,  E\right)$ 
coincides with the Holtsmark distribution corresponding to the concentration $n_i \left( \boldsymbol 0 \right) = n_{i0}$.

    Let us consider the behavior of the distribution function of the ionic microfield in the case of a Gaussian spatial distribution of electrons and ions:
\begin{equation}
    \label{Eq_8}
    n_{i,e} \left( \bm r \right) = \frac{N_{i,e}}{\left( 2 \pi \right)^{3/2} \sigma^3} \cdot exp \left[ -r^2/ 2 \sigma^2 \right]
\end{equation}
for small values of the ratio $r/\sigma$ and for large values of $N_i$.
For $r \ll \sigma$ (after replacing $\bm x = \bm r' - \bm r$):
\begin{widetext}
    \begin{multline}
        \label{Eq_9}
        C \left( \bm \mu ,  \bm r \right) = n \int d \bm x \cdot exp \left[ -\frac{ x^2 + 2 \bm r \bm x + r^2}{2 \sigma^2} \right]\cdot\left( 1 - exp \left[ -i e \frac{\boldsymbol \mu \bm x}{x^3} \right] \right) \approx \\
        C_0 \left( \mu \right) - i e \boldsymbol \mu n_{i0} \int \frac{\bm x}{x^3} d \bm x \cdot exp \left[ -\frac{x^2}{2 \sigma^2} \right] \frac{\bm r \bm x}{\sigma^2} \approx 
        C_0 \left( \mu \right) + i \boldsymbol \mu {\bm E}_i \left( \bm r \right),\\
        \text{ here }
        C_0 \left( \mu \right) = \left( E_0 \mu \right)^{3/2}  \left( \bm r \right) \text{ and } {\bm E}_{i,e}  \approx \pm \frac{4 \pi}{3} e n_{{i,e}0} \bm r.
    \end{multline}
    \text{Finally, substituting~(\ref{Eq_9}) and ${\bm E}_m = {\bm E}_e \left( \bm r \right) + {\bm E}_i \left( \bm r \right)$ into~(\ref{Eq_5}) we obtain:}
    \begin{multline}
        \label{Eq_10}
        P \left( \bm r, \bm E \right) \approx \frac{1}{\left( 2 \pi \right)^3} \int d \boldsymbol \mu \cdot exp \left[ i \boldsymbol \mu \Big( \bm E - {\bm E}_m \left( \bm r \right) \Big) \right] \cdot
        exp \left[ -\left( E_0 \mu \right)^{3/2} \right] \approx \\
        P_0 \left( \bm r ,  \bm E \right) - \frac{1}{\left( 2 \pi \right)^3} \int \left( i \boldsymbol \mu {\bm E}_m \left( \bm r \right) + \frac{1}{2} \mu^2 {E}_m^2 \right) \cdot 
        exp \left[ i \boldsymbol \mu \bm E \right] \cdot exp \left[ -\left( E_0 \mu \right)^{3/2} \right] d \boldsymbol \mu.
    \end{multline}
    \text{And for small $r$:}
    \begin{multline}
        \label{Eq_11}
        p \left( \beta ,  r/\sigma \right) = E_0\widetilde{p}(\bm r, E_0\beta) = p_o(\beta) - \frac{E_0^3E_m^2\beta^2}{2 \left( 2 \pi \right)^3} \cdot 
        \int \mu^2 exp \left[ i E_0 \boldsymbol \mu \boldsymbol \Omega \beta \right] \cdot exp \left[ -\left( E_0 \mu \right)^{3/2} \right] d \boldsymbol \mu \approx \\
        p_0(\beta) - \frac{2}{9\pi} \left( \frac{15}{4} \right)^{4/3} \cdot \left( \frac{n_{i0}-n_{e0}}{n_{i0}} \right)^2 \cdot \left( N_i^{1/3} \frac{r}{\sigma} \right)^2\beta
        \int \limits_{0}^{\infty} x^3 sin \left( \beta x \right) \cdot exp \left[ -x^{3/2} \right] dx
    \end{multline}
    \text{Here $E = E_0 \beta$ and $p_0 \left( \beta \right)$ is Holtsmark distribution.}
\end{widetext}
    The characteristic scale of the field strength for the distribution function at $\bm r = \boldsymbol 0$ corresponds to 
$E_0 = \left( 4/15 \right)^{2/3} e N_i^{2/3}/\sigma^2 = 2 \pi \left( 4/15 \right)^{2/3} e n_{i0}^{2/3}$. The characteristic scale of the field strength at the periphery is 
$0.27 e \left( N_i - N_e\right)/\sigma^2$ (the value of the solution of the equation $\bm \nabla \cdot \bm E = 4 \pi e \left( n_i - n_e \right)$ the point of maximum $r/\sigma = 1.37)$.Therefore,
the relative role of the macrofield and microfield in the formation of the line shift ($\sim E^2$) is determined by the ratio of $\left(N_i - N_e\right)^2$ to $N_i^{4/3}$.
For $N_i - N_e$, the following estimate holds: 
$\left( N_i - N_e\right)^2 \sim N_i k_{B} T_e \sigma/e^2 \sim N_i^{4/3} k_{B} T_e/e^2 n_{i0}^{1/3}$~\cite{B_Vikhrov_2,B_Killian}. Therefore, the condition 
for the predominance of the macrofield in the formation of the distribution function coincides with the condition of the weak coupling of the plasma $k_{B} T_e/e^2 {n}^{1/3} \gg 1$.
\begin{figure}
    \includegraphics[width=1\columnwidth]{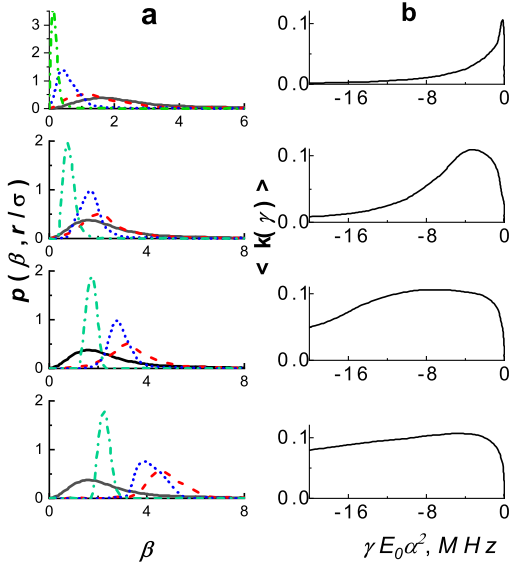}
    \caption{
                \label{Fig_2} (a) $p\left( \beta , r/\sigma \right), \left( \beta = E/E_0, E_0=0.2 V/cm \right)$ for $N_i = 10000$ and for four values of 
                $\left( N_i - N_e \right)/N_i$: $0.97$, $0.83$, $0.70$, $0.57$, which correspond to the experimental conditions of~\cite{B_Park}. From top to bottom: 
                $T_{e0}=1$~K, $62$~K, $134$~K, and $206$~K. Curves are shown for different values of $r/\sigma$: $0$ corresponds to the black solid line, $1$ corresponds to the red dashed line, $2$ corresponds to 
                the blue dotted line and $3$ corresponds to the green dash-dotted line. (b) Dimensionless absorption coefficient for the parameters corresponding to the experiment of~\cite{B_Park}. From top to bottom: 
                $T_{e0}=1$~K, $62$~K, $134$~K, and $206$~K.
            }
\end{figure}

    Fig.~\ref{Fig_1} shows he distribution functions $p \left( \beta, r/\sigma \right)$ calculated according to~(\ref{Eq_5}),~(\ref{Eq_6}) and~(\ref{Eq_7}) for two values of $N_i$  and for four values of 
$r/\sigma$. The values of $N_e/N_i$ on the upper graphs ($N_e/N_i$ = 0 which correspond to the maximum value of the macrofield), on the lower graphs ($N_e/N_i = 1$ 
(there is no macrofield)) and on the middle graphsare such that the influence of the microfield and macrofield is comparable in terms of magnitude: for $N_i = 1000$, $N_e/N_i = 0.6$; while 
for $N_i = 10000$, $N_e/N_i = 0.2$.

    The transition energy shift ($h \nu$) is proportional to the squared value of the field strength ($\nu_0 \to \nu_0 + \alpha E^2$), and the frequency dependence of the absorption coefficient is determined by the 
distribution function of the field strength.

For the absorption coefficient normalized to unity, we have:
\begin{multline}
        k \left( \gamma ,  \xi \right) = \int \limits_{0}^{\infty} \delta \left( \nu - \nu_0 - \alpha E_0^2 \beta^2 \right)\cdot p \left( \beta ,  \xi \right) d \beta = \\
        \frac{1}{2 \sqrt{\gamma}} p \left( \gamma , \xi \right), \text{ where } \gamma = \frac{\nu - \nu_0}{\alpha E_0^2} \text{ and }\xi = \frac{r}{\sigma}
\end{multline}
    In an experiment, the absorption coefficient averaged in space along the path of the probing beam is usually observed.
As applied to a beam passing through the center of a Gaussian distributed plasma, we have the following expression for the absorption coefficient normalized to unity and averaged over $\xi$:
\begin{equation}
    \langle k \left( \gamma \right) \rangle = \frac{2}{\sqrt{\pi \gamma}} \int \limits_{0}^{\infty} exp \left[  - \xi^2 / 2 \right]\cdot  p \left(  \sqrt{\gamma} ,  \xi \right) d \xi.
\end{equation}
Figure~\ref{Fig_2}(a) shows the calculation results for $p \left( \beta ,  r/\sigma \right)$ ($\beta = E/E_0$, $E_0 = 0.2$~V$/$cm), for $N_i = 10000$ and for four values of 
    $\left( N_i - N_e \right)/N_i$: $0.57$, $0.70$, $0.83$, $0.97$, which correspond to the experimental conditions of~\cite{B_Park} ($T_{e0} = 206$~K, $134$~K, $62$~K, and $1$~K).

    It can be seen that at small values of $N_e/N_i$, the macrofield has a strong influence on the distribution, and with $r/\sigma$ increasing, it first shifts towards larger values of $\beta$ then, 
after passing through the maximum of the macrofield at $r/\sigma = 1.37$, it returns to smaller values, demonstrating increasing similarity with the delta function. At values of $N_e/N_i$ close to unity,
the macrofield has little effect on the distribution function and it monotonously shifts towards smaller $r/\sigma$ when $\beta$ increases.

    Figure~\ref{Fig_2}(b) shows the values of the absorption coefficient for the parameters corresponding to the experiment of~\cite{B_Park}.
\begin{figure}
    \includegraphics[width=1\columnwidth]{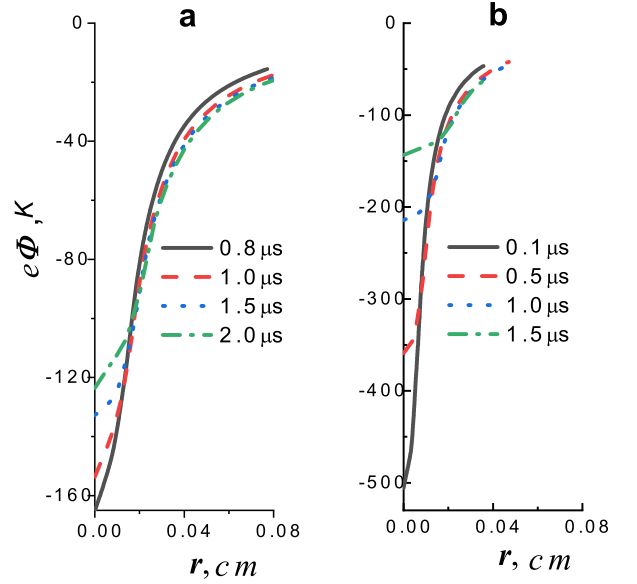}
    \caption{
                \label{Fig_3} Dependence of the potential of the well formed due to the charge imbalance in the Rb plasma cloud on the radius and expansion time for $N_i = 5000$. (a) $T_{e0}=25$~K; 
                (b) $T_{e0}=100$~K.
            }
\end{figure}
Using the data shown in Fig.~4 of~\cite{B_Park}, it is possible to establish the magnitude of the electric field that ensures the shift of the maximum of the $42p_{1/2}$ resonance line for $T_e=1$~K.
It amounts to $E \approx 0.2$~V$/$cm ($\alpha = -33$~MHz$/($V$/($cm$)^2)$. Since the contribution of the macrofield at this temperature can be considered the smallest, we can put $E \approx E_0$. Using the obtained value 
of $E_0$, it is possible to determine the shifts of the absorption coefficient maximum for other temperatures and to compare them with experiment. The table shows the calculated and experimental values 
(obtained from Fig.~4 in~\cite{B_Park}).
\begin{table}
    \caption{\label{T1}Shift $\Delta \nu$ of the maximum of the $42p_{1/2}$ spectral line as a function of $T_e$.}
    \begin{tabular}{ccccc}
        \hline
        \hline
        $T_e$,~K                       & 		$1$&                            	$62$& 			$134$&			$206$\\
        $\Delta \nu_{exp}$,~MHz&			$1.3 \pm 0.5$&		$4 \pm 0.8$&     		$10 \pm 2.0$&		$13 \pm 2.5$\\
        $\Delta \nu_{calc}$,~MHz&			$0.8 \pm 0.2$&		$4 \pm 0.7$&		$8 \pm 1.5$&		$11 \pm 2.0$\\
        \hline
        \hline
    \end{tabular}
\end{table}
    According to the authors of~\cite{B_Park}, the error in determining $\Delta \nu$ in the experiment is 10\%. However, when using graphic material, this error becomes larger. We estimated the error of the numerical 
calculation at 20\%. Nevertheless, it is possible to speak of satisfactory agreement between the experimental and theoretical data.

    In addition, it is necessary to discuss some questions which are important for the use of the results obtained.

Firstly, the distribution function considered above for the ionic microfield in a plasma cloud with a Gaussian density distribution in the case of non-interacting charges is a zero approximation in the analysis of a real
plasma. In our works ~\cite{B_Bobrov, B_Bronin}, we calculated the distribution functions of the ionic microfield for a two-component homogeneous plasma as a function of the nonideality parameter. These results
can use for specific strongly coupled parameters. They can also use taking into account the change in density and temperature at each instant of plasma expansion. All these recommendations apply to the distribution 
function depending on the radius as well. Moreover, for small strongly coupled parameters and for large radii, the use of the Holtsmark distribution is justified.

    Secondly, in this paper, for calculating the distribution function of the electric field, we used an approximation where the ion density is distributed according to the Gauss function. As the plasma cloud expands 
this distribution is valid only for $\sim 1$ $~\mu$s. Figure~\ref{Fig_3} shows an example of the results of MD simulation of macrofield formation in a rubidium plasma cloud at $N_i = 5000$ for two values of 
imbalance corresponding to $T_{e0}=25$~K and $T_{e0}=100$~K. Our calculations show that during the formation of the potential well, the ion density distribution coincides with the Gaussian distribution.
In Figs.~\ref{Fig_3}(a) and~\ref{Fig_3}(b), this corresponds to the potential curves for $t =0.8$ and $0.2$~$\mu$s, respectively. After this time lapse, a peak in the ion density is formed at the cloud boundary, which 
peak results in the formation of an ion wave. This must be taken into account when calculating the distribution function for long expansion times.

    In conclusion, we present a general expression for the probability distribution function of electric field in a spatially inhomogeneous non-neutral plasma. This expression takes into account the ionic microfield and 
the macrofield arising due to the charge imbalance depending on the size of the plasma cloud. The results of numerical calculation of the distribution of electric field in the case of non-interacting particles are 
presented. A method is proposed for taking into account the influence of the interaction between charges on the distribution of the ionic microfield in a spatially inhomogeneous non-neutral plasma. As the charge imbalance 
increases, the effect of a sharp increase in the distribution function for large field values is observed. Good agreement between the calculation of the shift of the spectral line and the experiment is obtained. 
The results obtained are of crucial importance for diagnosing the properties of such plasma, both low-density plasma and strongly coupled plasma, and can be widely used in various applications.

\end{document}